\documentclass{ptptex}
\usepackage{graphicx}%

\usepackage{amsmath}%

\usepackage{amsfonts}%
\usepackage{amssymb}

\usepackage{bm}
\begin{document}
\markboth{T. Boness, M.M.A. Stocklin and T.S. Monteiro}{Quantum chaos with
spin-chains in pulsed magnetic fields}
\title{Quantum chaos with spin-chains in pulsed 
magnetic fields }
\author{T. Boness,  M.M.A. Stocklin and T.S. Monteiro}
\inst{Department of Physics and Astronomy, University College London, Gower Street, London WC1E 6BT, United Kingdom}
%
%
\notypesetlogo

\abst{Recently it was found that the dynamics in a Heisenberg 
spin-chain subjected to a sequence of periodic pulses from an external,
parabolic, magnetic field can have a close correspondence with 
 the quantum kicked rotor (QKR). The QKR is a key paradigm
of quantum chaos; it has as its classical limit the well-known Standard Map.
It was found that a single spin excitation could be converted into
a pair of non-dispersive, counter-propagating spin coherent states
equivalent to the accelerator modes of the Standard Map.
 Here we consider how other types of quantum chaotic systems such as a
double-kicked quantum rotor or a quantum rotor with a double-well potential
 might be realized with spin chains; we discuss
the possibilities regarding  manipulation of the one-magnon spin waves. }
 
\maketitle
\newcommand{\ket}[1][\psi]{| #1 \rangle}
\newcommand{\bra}[1][\psi]{\langle #1 |}

\section{Introduction}

It has been shown that the quantum properties of certain many
body systems  may be analyzed by considering the
dynamics of an analogous `image' one-body  system \cite{Prosen}. In the case
of certain types of spin-chains, there can be one-body 
 image systems with a well-defined classical limit, which can be chaotic or integrable
\cite{Prosen2}.

In another context, there has also been considerable interest in the dynamics of 
quantum spin-chains because of their potential for quantum information
 applications. Quantum state transfer is one example: the ability to transfer
a qubit, or some arbitrary quantum state, with high fidelity along a spin-chain
has been addressed in several works; in \cite{Bose} quantum state transmission
of a single spin-flip  along a ferromagnetic Heisenberg chain was investigated. 
In a subsequent work \cite{Shi} it was shown that such
a chain, in the presence of an external, static, parabolic magnetic field,
can give perfect transmission if the initial spin state is a specific coherent state.

Recently \cite{Boness} a close correspondence between the unitary time evolution
of the ferromagnetic Heisenberg spin-chain -for a single spin-excitation- and
that of the Quantum Kicked Rotor (QKR) was noted. To complete this
correspondence, a periodically-pulsed parabolic magnetic field must also be
applied.  The QKR is a leading paradigm of quantum chaos \cite{Casati} which has
been extensively studied theoretically. Its classical counterpart is one of the
best-known `textbook' examples of Hamiltonian chaos, the Standard Map
\cite{Ott}.  The QKR has also been investigated experimentally- since it was
shown by a cold atom group in Texas that a realization with cold atoms with
pulsed standing waves of light is possible \cite{Raizen}. A number of  other
aspects of this system in its quantum chaotic regime  were also
studied by this and several other experimental cold atom groups
\cite{expt,Jones}.

In \cite{Boness} it was found  that with the pulsed parabolic field, 
one can employ certain  expressions found 
for the QKR and the Standard Map to describe the entanglement properties.
It means also that the system exhibits not only generic
forms of quantum chaotic behaviour in the spin-chain 
like exponential localization (analogous to Anderson localization
 seen in disordered metals) but also generate phenomena such
as `accelerator modes' which are a feature specific to the Standard Map alone.

In \S\ref{sec:khsc} we review the QKR one-body image of the ferromagnetic
Heisenberg spin-chain.  In \S3 we show that the double-kicked rotor (QKR-2), a
system which has been the subject of an experimental study \cite{Jones} and has
been found to have rather different dynamics from the QKR, also has a spin-chain
analogue.  We propose that rather than implementing the QKR-2 which is a KAM
dynamical system, it may be easier to implement a closely related random map.
In \S4 we show that  double-well kicked rotors may be implemented with
spin-ladders or ferromagnets with next-to-nearest neighbor exchange interaction.
To conclude, we discuss the possibility of investigating the corresponding
antiferromagnetic dynamics and we discuss the potential uses of the quantum
chaos as a means to manipulate spin waves.

\section{Quantum kicked rotors with Heisenberg ferromagnetic spin-chains}
\label{sec:khsc}
The Heisenberg spin-chain has the well-studied Hamiltonian:
\begin{eqnarray} 
H_{hc} = -J \sum_{n} \sigma^n \cdot \sigma^{n+1} -\sum_{n} B \sigma_z^n.
\label{eq1}
\end{eqnarray}
For a ferromagnet, $J >0$.
We consider the effect of applying a periodic sequence of short pulses from a
 parabolic magnetic field. We can model these by a series of $\delta-$kicks. Note that in
equivalent QKR atomic experiments the pulses will, of course, have finite
duration. The combined  Hamiltonian is:
\begin{eqnarray}
{\bf H}= H_{hc} + \sum_{n=1}^{N} \frac{B_Q}{2}(n-n_0)^2 \sigma_z^n \sum_j \delta(t-jT_0)
\label{eq2}
\end{eqnarray}
where $T_0$ is the period of the pulses; $B_Q$ is the amplitude 
of the applied parabolic magnetic field; the length of the chain, 
$N \gtrsim 100$ in the present work. In \cite{Boness} this effect was
investigated for a single excitation on an open chain.  Here we consider the
simpler case of a chain with periodic boundary conditions (eg. a ring).
However, for long chains, the dynamics we explore are not sensitive to these
boundary conditions.

Since $[{\bf H}, S_z] = 0$, where $ S_z$ is the total spin component in the $z$
direction, if we prepare an initial quantum state comprising all spins pointing
`up' and a single spin `down' at some arbitrary site $\psi(t=0) = \ket[{\bf
s_0}]$,  one may, for all $t$,  obtain $\psi(t)$ in terms of a basis of states
$\ket[ {\bf s} ] $, which have  a spin-down at a single site $s$ on the chain but
all other spins up (along ${\hat z}$). The eigenstates of $H_{hc}$ in this
basis distribute the spin-flip periodically along the chain:
\begin{equation}
    \ket[\tilde m] =  \frac{1}{\sqrt{N}} \sum_{j=1}^{N} e^{i \ j k_m} \  \ket[
{\bf j} ].
\label{eq3}
\end{equation}
These states represent spin waves (magnons) with wavenumber $k_m =
2m\pi/N$, $2m \in (-N,N]$ and may be obtained using the Bethe ansatz.
The corresponding eigenenergies are given by the  one-magnon dispersion relation
for a ferromagnet:
\begin{equation}
E_m - E_0 = J(1- \cos k_m) + B,
\label{eq4}
\end{equation}
where, $E_0$ is the ground state energy $-JN/4-NB/2$.

An analytical form for the time-evolution
operator $U_{hc}(t,0) = \exp \{ -\frac{i}{\hbar}H_{hc}t \}$ may be given,
in the single-excitation basis, as a matrix of elements:
\begin{equation}
U^{hc}_{rs}(T_0)= e^{-i \frac{B_Q}{2}(r-n_0)^2} \ \frac{1}{N}\sum_m  e^{i {
(r-s) k_m + i JT_0 \cos k_m} }.
\label{eq5}
\end{equation}
Here, we disregard the overall phase due to $E_0$ and the uniform static field $B$
(or formally set $2BT_0=2 \pi$). In an actual realization, the external field is 
significant as it introduces a gap with the fully aligned ground state: it does not
modify the dynamics of interest here.

In comparison, the QKR has Hamiltonian:
\begin{eqnarray}
H=\frac{P^2}{2} - K \cos x \sum_n \delta(t-nT).
\label{eq6}
\end{eqnarray}
Free evolution of the particles is followed by a short kick from a sinusoidal potential.
 $K$ (stochasticity parameter of the Standard Map) is the kick-strength; it is related to 
the intensity of the optical lattice in atomic experiments.
In a basis of momentum states $\ket[l]$, where $p=l\hbar$, the QKR
unitary evolution operator has matrix elements:

\begin{equation}
U^{QKR}_{nl}(T)= e^{-i \frac{T\hbar}{2} l^2} \ \frac{1}{2 \pi}
\int_{-\pi}^{\pi}
e^{i {(n-l)x +i \frac{K}{\hbar} \cos x} } \ dx .
\label{eq7}
\end{equation}

From (\ref{eq5}) it is evident that the `free evolution' part of $U^{hc}$ due
the static spin-exchange interaction is simply a discretized version of the kick
part of $U^{QKR}$; while, in turn, the `kick' provided by the parabolic magnetic
field is equivalent to the free-evolution of the QKR. The magnon wavenumber maps
onto position for the QKR, ie $k_m \to x$; similarly, the spin
site maps onto momentum, so $ s \to l$.  Since, for the QKR, we rescale time
units so $T=1$, we see that in (\ref{eq5}), $B_Q \to \hbar$ ie the parabolic
field can be considered an effective value of $\hbar$ in the QKR image system.
Finally, the classical stochasticity parameter, $K= JT_0 B_Q$.

\begin{figure}[t]
\begin{center}
\includegraphics*[angle=0,width=3.5in]{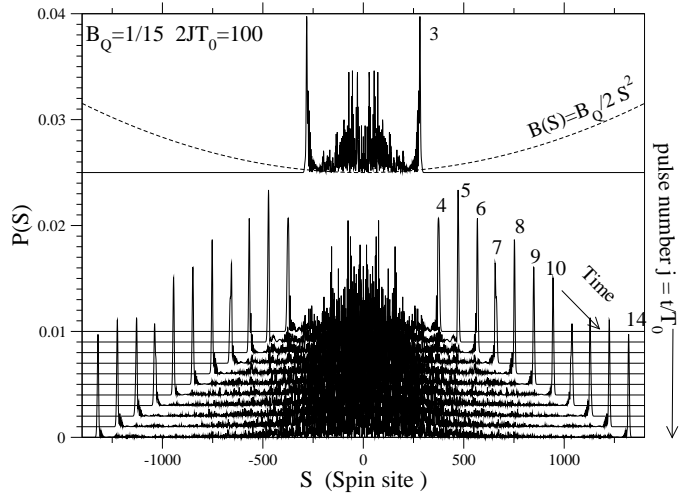}
\end{center}
\caption{Effect of accelerator modes in a Heisenberg spin-chain. $P(s)$ represents
the probability of finding the excitation at site $s$.
 The accelerator modes are the `spikes' at the leading edge of the distribution.
They correspond to a counter-propagating pair of coherent states ejected from the centre.  
 We take $B_Q=1/15$ and $JT_0=100$. The upper line 
is at $t=3T_0$; the lower curves correspond to consecutive periods $jT_0$
with $j=4,5,6...$ as numbered. The dotted line indicates the form of
the parabolic field (scaled by a constant factor)
 which is pulsed on/off every period  at times $t=jT_0$.  
The accelerator modes represent over $25\%$ of the total probability;
they advance an equal distance (shown below to be $ 2\pi/B_Q \simeq 94$ spin sites)
 each period, and after just 3 pulses are well separated from the central, `chaotic'
remnant.}
\label{Fig1}
\end{figure}

Equations (\ref{eq5}) and (\ref{eq7}) are sufficiently close that the behavior of the QKR
maps onto the Heisenberg spin chain at even the detailed level.
We see, not only, exponential localization for strong $K$, but also, for values of
$K$ slightly above $K=2\pi$, we see accelerator modes. Accelerator modes are a
peculiarity of the standard map: deep into the full chaotic regime, small transporting
stable islands re-appear and cause ballistic rather than diffusive transport
over part of phase-space. This is illustrated in  Fig.{\ref{Fig1}}.
 We show, for $K \simeq 6.6$, the effect of repeated application of 
the unitary matrix of (\ref{eq5})
on a state initialized with a spin flipped `up' on a site for 
which  $ s \simeq 2\pi M/B_Q$ where $M=0,1,2,...$ which is close to an accelerator island.
 In Fig.{\ref{Fig1}} we took simply $M=0$ and  $\psi(t=0) = \ket[n_0]$, ie a site near the center
of the chain.
 The spin-amplitude spreads out into a
very irregular `chaotic' distribution around the site $n_0$. But most strikingly, we can
see a pair of  counter-propagating spikes, `hopping' around $2\pi/B_Q \simeq 94$ spin-sites each 
 consecutive period. 

Away from parameters where there are accelerator modes, but still in the
chaotic regime $K \gtrsim 4$, the QKR exhibits an analogue of Anderson localization
in disordered metals; for the QKR, this means exponential localization
of momentum. For the Heisenberg spin-chain one can straightforwardly infer
(and verify numerically \cite{Boness}) that a single excitation will, after a timescale
$t > t^* \sim (JT_0)^2$ give rise to an exponential spin probability distribution:
\begin{equation}
    P(s) \sim \frac{2}{L} \exp \{ -2|s-s_0|/L \}.
    \label{eqPLOC}
\end{equation}
where $L \simeq (JT_0)^2/4$.
Note that the variance of the spin distribution $<(s-s_0)^2> \simeq L^2$
is independent of the initial state $\ket[ s_0]$. Similarly,
the asymptotic variance of the QKR momentum distribution 
$<(p-P_0)^2> \simeq K^2/\hbar $
 is independent of the initial momentum of the atoms, since at 
$t=0$, typically, the momenta of the cold atom cloud is 
close to some value $P_0$. Typically $N(p) \sim \exp{(p-P_0)^2/\sigma^2}$,
where $\sigma \simeq 4$ in the usual units depends on the initial
temperature of the cold atoms .

\section{Double-kicked rotors (QKR-2) with spin-chains}

In \cite{Jones} the double-kicked rotor (a variant of the QKR) was investigated
experimentally with cold atoms: a cloud of Cesium atoms was subjected to pairs
of closely spaced pulses from standing waves of light.  The corresponding
Hamiltonian is:
\begin{eqnarray}
H=\frac{P^2}{2} - K \cos x \left[\sum_n \delta(t-nT)+ \delta(t-nT+\epsilon) \right]
\label{eq9}
\end{eqnarray}
with period $T$, kick strength $K$ and $\epsilon \ll T$ the time between each kick in the
pair.

\begin{figure}[b]
\begin{center}
\includegraphics[width=2.5in,height=1.9in]{Fig2.eps} 
\hspace{0.5in}
\includegraphics[width=2in,height=2.3in]{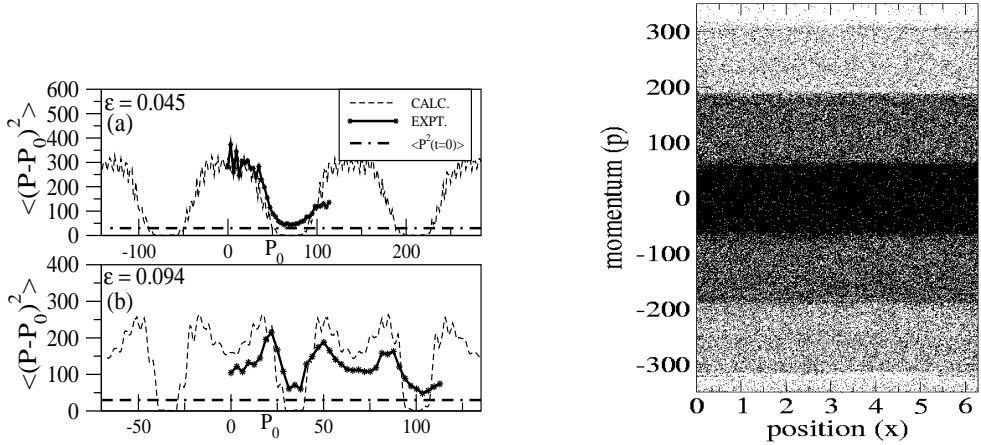}
\end{center}
\caption{{\bf Left:} Experimental momentum variances for the 2-QKR, obtained
with cold atoms. \cite{Jones} The figure shows that at trapping
 sites $P_0 \simeq (2m+1)\pi/\epsilon$ there is little
spreading, while in between, $< (p-P_0)^2> \sim (\pi/\epsilon)^2$.  
{\bf Right:} Surface of section plot for the $2\delta$-KP, $K_\epsilon=0.35$,
 for arbitrary, large $\tau_\epsilon \to \infty $,
illustrating  the classical dynamics of the `one-body image'
of the double pulsed spin chain . The cellular structure is evident:
phase space is divided into regions of fast diffusion
separated by porous boundaries, ie narrow trapping regions.
For the equivalent spin-chain, this would 
correspond to spin sites $s$ for which $s B_\epsilon \simeq \pm(2m+1)\pi$.
 where $m=0,1,2,..$. Spin excitations prepared at these sites
 would remain highly localized, while those prepared in between
trapped sites would spread over one whole cell.} 
\label{Fig2}
\end{figure}

The variances of the asymptotic (long time) distribution in the QKR-2 depend
strongly on the initial state as shown in Fig.{\ref{Fig2}}.  The results display
`momentum trapping' regions where atoms absorb little energy if
$P_0\simeq(2m+1)\pi/\epsilon$.  Here the variances remain close to the initial
values (which depend on the temperature -of order $\mu$K- of the cloud. In
between these trapping regions, the typical variances  $\sim  \pi/\epsilon$ are,
unlike the QKR,  independent of both the kick strength and of $\hbar$.

Fig.{\ref{Fig2}} also shows a typical surface of section. Classical
phase-space shows a cellular structure, with the cell-boundaries
corresponding to the momentum trapping regions. In \cite{Creff} properties
of the QKR-2 were investigated. For $T \simeq 1-2$,
the trapping regions correspond to thin mixed-phase-space regions
filled with islands and cantori. As $T \to \infty$ all islands
disappear and a non-KAM limit, with no islands, is approached.
But a remarkable feature of this system's dynamics is that the trapping
effects remain as strong, even as $T \to \infty$.

The classical map for the $2\delta$-KP is a simple
extension of the Standard Map:
\begin{eqnarray}
p_{N+1}=p_N - K\sin x_N; &\quad & p_{N+2}=p_{N+1} - K\sin x_{N+1}
\nonumber \\
x_{N+1}=x_N + p_{N+1}\epsilon; &\quad & x_{N+2}=x_{N+1} + p_{N+2}\tau
\nonumber \\
\label{eq10}
\end{eqnarray}
where $\tau=T-\epsilon $ is the time interval between the kick-pairs.
 Starting from the map (\ref{eq10}) with $N=0$ we
re-scale all variables: $p^{\epsilon} = p\epsilon$, $K_{\epsilon}
= K \epsilon$ and $\tau_{\epsilon}= \tau/\epsilon \gg 1$ to obtain
a re-scaled map for which the momentum `cells' are of width $2\pi$.  While it is
usual for the kicked rotor to be re-scaled in terms of $T$, the appropriate
scaling here is in terms of $\epsilon$:
\begin{eqnarray}
p^{\epsilon}_1=p^{\epsilon}_0 - K_{\epsilon}\sin x_0; &\quad &
p^{\epsilon}_2=p^{\epsilon}_1 - K_{\epsilon}\sin
x_1 \nonumber \\
x_1=x_0 + p^{\epsilon}_1; &\quad & x_2=x_1 +
p^{\epsilon}_2\tau_{\epsilon}.  \label{eqn11}
\end{eqnarray}

In the atomic experimental and theoretical studies,
\cite{Jones,Creff} the regime $\tau_{\epsilon}= 20-100$  was investigated.
However, it was found in \cite{Mischa}  that diffusion correlations
which determine the trapping depend only on $K_\epsilon$; the dynamics
is completely insensitive to $\tau$ of  $\tau_{\epsilon} \gtrsim 50$.
In effect, very similar dynamics is obtained if we replace $x_0$ in 
(\ref{eqn11}) by a random number in the interval $[0,2\pi]$. 
The classical diffusion/trapping depends only on correlations
between kicks in each pair; consecutive kick pairs are completely
uncorrelated for $\tau_{\epsilon} \gtrsim 50$.
We can exploit this useful feature to facilitate an implementation of the QKR-2
with a Heisenberg spin-chain.

The time evolution operator for this system may be written:
\begin{eqnarray}
\hat{U}_{QKR-2}^{\epsilon} = e^{\left[-i \frac{l^2 \hbar\tau}{2}\right]}
e^{\left[i\frac{K}{\hbar} \cos x\right]} 
e^{\left[-i \frac{l^2 \hbar \epsilon}{2}\right]} \
e^{\left[i\frac{K}{\hbar} \cos x\right]}.
\label{eq11}
\end{eqnarray}
It is easy to see that $U$ is
invariant if the products $K_\epsilon= K\epsilon$ and
$\hbar_\epsilon= \hbar\epsilon$
are kept constant; while the free propagator
$U_l^{free}=e^{-i \frac{l^2\hbar \tau}{2}}$
simply contributes a near-random phase. Provided that
$l^2T \hbar \gg 2\pi$,
the results are quite insensitive to the magnitude of $(T-\epsilon)\hbar$.

Now, for the image Heisenberg spin chain of the QKR-2, the quantum map corresponding to
(\ref{eq11}) is:
 \begin{eqnarray}
\hat{U}_{hc}^{\epsilon} = 
e^{\left[-i \sum_{n=1}^{N} \frac{B_\tau}{2}(n-n_0)^2 \sigma_z^n \right]}
e^{\left[-i{H_{hc}(T_0)}\right]} \nonumber \\
e^{\left[-i  \sum_{n=1}^{N} \frac{B_\epsilon}{2}(n-n_0)^2 \sigma_z^n \right]}
e^{\left[-i{H_{hc}(T_0)}\right]}.
\label{eq12}
\end{eqnarray}
Hence evolution under the Heisenberg Hamiltonian ${H_{hc}}$ for a period $T_0$
is followed by a short impulse from a parabolic magnetic field; however now the
strength of the parabolic field alternates in value between a strong field
$B_\tau$ and a weak field $B_\epsilon$ . We see that the ratio
 between the fields corresponds to the parameter
$\tau_\epsilon$ of the QKR-2, so $B_\tau/B_\epsilon \gg 1 \equiv \tau_\epsilon$.

If we alternate the field strengths, we find that the image spin-chain will
show all the dynamics of the QKR-2: a cellular structure for the phase space,
with most initial states evolving to a final state with variance of the spin-probability
$<(s-s_0)^2> \simeq (\pi/B_\epsilon)^2$. In other words, for $JT_0 \gg 1$ the single excitation
spreads out more or less uniformly over a segment of the chain of width $\simeq \pi/ B_\epsilon$
but remains trapped within this segment. States prepared at the edges of the
cells would remain highly localized over just a few spin sites.

\begin{figure}[t]
\begin{center}
\includegraphics[height=2in,width=4in]{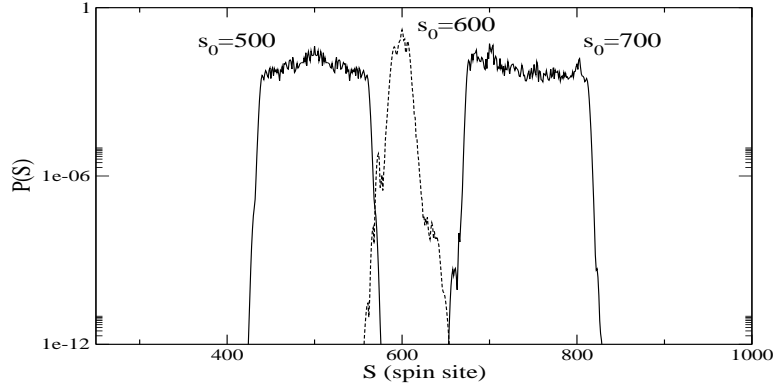}
\end{center}
\caption{Double-kicked spin chain with the strong parabolic field replaced by a
randomly varying field in time. This effectively removes correlations between
`kick pairs' in the evolution and causes spin excitations to remain trapped
within one segment of the chain, or between segments if prepared in a trapping
region. The figure shows two adjacent cells for $B_{\epsilon}=0.025$ and
$JT_0=7$, leading to effective $K_{\epsilon}=0.175$; $s_0$ denotes the starting
position of the excitation along the chain. Excitations that start at the center
of the chain ($s_0=500$) spread over one whole cell but no further. An
excitation starting off-center in the neighboring cell at $s_0=700$, spreads
over that cell, but not beyond the trapping region at $s_0\simeq 600$.
Excitations prepared near the center of the trapping region remain highly
localized over about 30-40 spin sites. Near the edges excitations can `escape',
but the probability is very low as shown.}
\label{Fig.4}
\end{figure}

However, as it is reasonable to suppose that a weak parabolic magnetic field might
pose less of a technical challenge than a strong parabolic field, we note that
the strong field can be simply replaced by a random phase, ie in (\ref{eq12}) we replace
 the term $ \exp {\left[-i \sum_{n=1}^{N} \frac{B_\tau}{2}(n-n_0)^2 \sigma_z^n \right]}
  = \exp {-i (\beta_n \sigma_z^n)}$ where 
$ \beta_n$ is a  magnetic field of randomly varying magnitude, along the $z$ direction.
In Fig.{\ref{Fig.4}} we show spin probability distributions obtained for the double spin
chain where we take the random field $\beta_n$ to be uniformly distributed in the interval
$[0, 2\pi]$. This corresponds to the dynamics of the (non-KAM) $\tau_\epsilon
\to \infty$ limit.

\section{Double-well kicked rotors and anti-ferromagnets}

Spin-chains in general can have longer ranged exchanged interactions beyond the
nearest-neighbor type of (\ref{eq1}). In particular, next-to-nearest neighbor
(NNN) interactions are well-known and are a feature of spin ladders.
\cite{Majumdar}.
\begin{eqnarray} 
H_{hc} =\sum_n\left( -{J_1} \sigma^n \cdot \sigma^{n+1}-{J_2}\sigma^n \cdot
\sigma^{n+2}\right) -\sum_{n} B \sigma_z^n
\label{eq13}
\end{eqnarray}

We restrict ourselves now to the case where the leading term is ferromagnetic ($J_1 >0$) but allow
the NNN term to have either a ferromagnetic ($J_2>0$) or antiferromagnetic
($J_2<0$) form.
While the latter can lead to magnetic frustration and become gapless, we do not concern ourselves
with that parameter range.

A simple Bethe-ansatz treatment of (\ref{eq13}) will obtain similar eigenstates
to (\ref{eq3}) but a modified
dispersion relation:
\begin{equation}
E_m - E_0 = (J_1 + J_2 - J_1 \cos k_m - J_2\cos 2k_m ) + B
\label{eq14}
\end{equation}
where $E_0 = -(J_1 + J_2)N/4-BN/2$ for the spin ladder.
Hence this system maps onto a kicked rotor with a `double-well' potential:
\begin{eqnarray}
H=\frac{P^2}{2} - (K_1 \cos x + K_2 \cos 2x) \sum_n \delta(t-nT),
\label{eq15}
\end{eqnarray}
where $K_{1,2}$ take the same signs as $J_{1,2}$. This type of potential has slightly different dynamics from the QKR;
if the kicking period (or for the spin-chain, the quadratic field) is varied, it can even give rise to a type
of Hamiltonian ratchet \cite{Mont}.

However, as a potentially useful means of manipulating spin-waves, it is worth
considering the regimes with low kick strength (small $K_1, K_2$) where the
classical dynamics has large stable islands. We recall that, in the absence of
the parabolic field, the Heisenberg ferromagnet (one-magnon) dynamics is
equivalent to the resonant QKR: quantum states are all fully delocalized and a
single spin flip will spread over the whole spin-chain.  However, kicking by a
weak field will generate large stable islands where spin-waves are trapped. The
location of these islands will determine whether spin excitations give rise
to low or high energy magnons -- low or high $k_m$.

\begin{figure}[htb]
\includegraphics[height=3in]{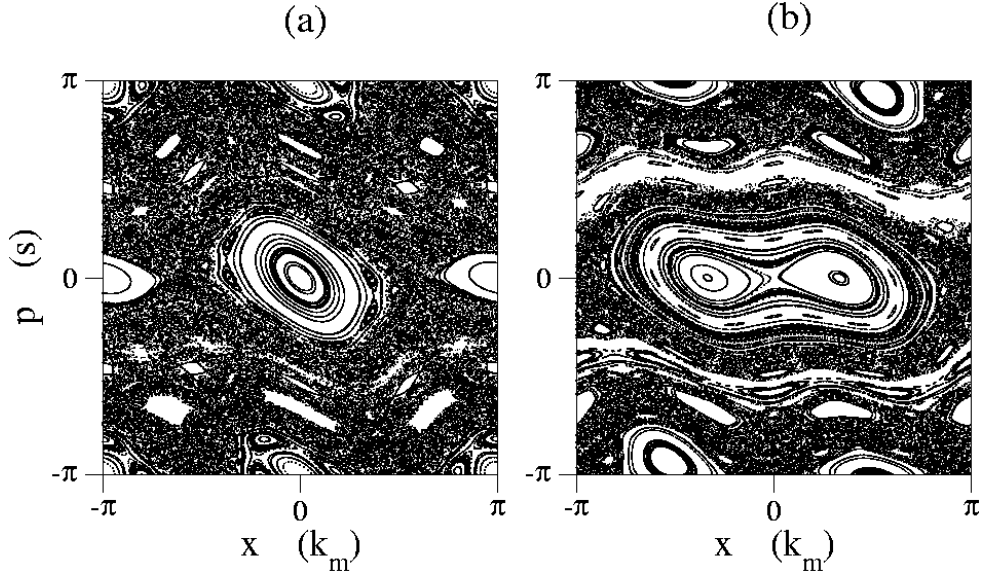}
\caption{Surface of section plots for the double-well kicked rotor with $K_1 =
0.35$ and $K_2 = \pm K_1$.  Stable islands can be used to influence the time
evolution of spin waves in a spin ladder with a ferromagnetic NN interaction.
(a) A ferromagnetic NNN interaction corresponding to $K_2=K_1$ will mean that a
spin-wave packet excited at low $k_m$ will remain confined at
low energies; while in (b), an antiferromagnetic NNN interaction ($K_2=-K_1$) will
favor intermediate $k_m$.
} \label{Fig5}
\end{figure}

For simplicity, we consider only the case $J_1= J_2 > 0$ as well as $J_1= -J_2 >
0$; we note, however, that it may be possible to adjust the
relative magnitude of $J_1$ and $J_2$ (possibly by varying the distance between
the adjacent spin-chains).  Fig.{\ref{Fig5}} shows the classical surfaces of
section for the ferromagnetic NNN and antiferromagnetic NNN respectively.

Finally, we consider briefly the case of an antiferromagnetic nearest-neighbor interaction.
The Bethe ansatz treatment obtains a one-magnon dispersion relation
for the spin waves $E_m =J |\sin k_m|$ , hence for $k \sim 0$ the observed
antiferromagnetic dispersion relation takes a linear $E_m \sim J |k_m| $ form.
This would make the equivalent kicked rotor dynamics somewhat analogous to
classical maps with discontinuities such as the tent-map or saw-tooth map.
Unfortunately, for usual antiferromagnets, these states are gapless. The 
$E_m =J |\sin k_m|$ relation represents a lower bound to a continuum of excitations.
It may be that an NNN type interaction could introduce a `gap'
between the linear dispersion relation and the continuum. Nevertheless, 
the prospect of a practical realization of a type of `saw-tooth map' using spin chains 
would represent a much harder technical challenge.

\section{Conclusion}

If parabolic fields can be generated one can show that the magnetic field strengths required
are not unreasonable. For example, we assume the parabolic field ranges from
 $B =0 -10^{-6}$  across a spin chain of $10^4$ sites
 ($B=10^{-6}$ atomic units corresponds to 0.47  Tesla, a modest laboratory field).
Hence $B_Q \sim 10^{-14}$. Since we require $N^2 B_Q \delta t \gg 1$, where
$\delta t$ is the duration of the short magnetic pulse, we need $\delta t \sim 10^{8}-10^{9}$ au
ie $\delta t \sim 1- 10$ nanoseconds. We note that picosecond pulses are possible
in current spin-wave experiments. We also require $2JT_0 \gg 1$, but for the
`split-operator' approximation in (\ref{eq5}), we implicitly assume $2J \delta t \ll 1$.
Since $J \sim 1GHz$ for ordinary Heisenberg ferromagnets, this not unreasonable,
if $T_0 \sim 1 \mu sec$, though it 
would be better to have a smaller value of $J \sim 0.1 GHz$. The smaller $J$ is,
the smaller $B_Q$ can be (and the longer the duration of the pulses). 
For the double-kicked rotor, provided it
is possible to introduce a random phase, the parabolic field could be
$100-1000$ times smaller.

We conclude that if parabolic magnetic fields can be applied to
quite common ferromagnetic spin chains, the classical chaotic 
dynamics of the QKR  would suggest a number of 
new ways of manipulating the transport of the spin waves.


\begin{thebibliography}{99}
\bibitem{Prosen} T.Prosen, Phys.\ Rev.\ Lett.\ {\bf 80} (1998), 1808.
\bibitem{Prosen2}T.Prosen, Phys.\ Rev.\ E {\bf 60} (1999), 1658; Phys.\ Rev.\ E {\bf
65} (2002), 036208.
\bibitem{Bose}S. Bose, Phys.\ Rev.\ Lett.\ {\bf 91} (2003), 207901.
\bibitem{Shi} T. Shi, Ying Li, Z. Song, and C. P. Sun, Phys.\ Rev.\ A {\bf 71}
(2005) 032309.
\bibitem{Boness} T. Boness , S. Bose and T.S. Monteiro, Phys.\ Rev.\ Lett.\ {\bf
96} (2006), 187201.
\bibitem{Casati}  G. Casati,  B.V. Chirikov, F.M. Izraelev, and J. Ford in
``Lecture notes in Physics", Springer, Berlin {\bf 93} (1979), 334;
 S. Fishman, D.R. Grempel, R.E. Prange, Phys.\ Rev.\ Lett.\ {\bf 49} (1982), 509.
\bibitem{Ott}E. Ott, `Chaos in dynamical systems', Cambridge University Press  (1993).

\bibitem{Raizen} F.L. Moore, J.C. Robinson, C.F. Bharucha, B. Sundaram, and M.G.
Raizen, Phys.\ Rev.\ Lett.\ {\bf 75} (1995), 4598.
\bibitem{expt}H. Ammann, R. Gray, I. Shvarchuck, N. Christensen,
 Phys.\ Rev.\ Lett.\ {\bf 80} (1998), 4111; B.G. Klappauf, W.H. Oskay, D.A. Steck, and M.G.Raizen,
 Phys.\ Rev.\ Lett.\ {\bf 81} (1998), 4044; M.K. Oberthaler, R.M. Godun, M.B. d'Arcy, G.S. Summy, K. Burnett,
 Phys.\ Rev.\ Lett.\ {\bf 83} (1999), 4447; P. Szriftgiser, J. Ringot, D. Delande, J.C. Garreau,
 Phys.\ Rev.\ Lett.\ {\bf 89} (2002), 224101.
\bibitem{Jones}P.H. Jones, M.M. Stocklin, G. Hur, T.S. Monteiro, Phys.\ Rev.\
Lett.\ {\bf 93} (2004), 223002.
\bibitem{Creff} C.E. Creffield, G. Hur, and T.S. Monteiro,
Phys.\ Rev.\ Lett.\ {\bf 96} (2006), 024103; C. Creffield, S. Fishman and T.S. Monteiro,
Phys.\ Rev.\ E {\bf 73} (2006), 066202.
\bibitem{Mischa} M.M.A. Stocklin and T.S. Monteiro, Phys.\ Rev.\ E {\bf 74}
(2006), 026210.
\bibitem{Majumdar} 
C.K. Majumdar and D.K. Ghosh, J.\ Math.\ Phys.\ {\bf 10} (1969), 1388;
C.K. Majumdar, J.\ Math.\ Phys.\ {\bf 10} (1969), 177.
\bibitem{Mont}T.S. Monteiro, P.A. Dando, N. Hutchings, M. Isherwood, Phys.\
Rev.\ Lett.\ {\bf 89} (2002), 194102.

\end{thebibliography}
\end{document}